\documentclass[a4paper,11pt]{article}
\pdfoutput=1 
\usepackage{jinstpub} 

\usepackage{float}
\usepackage{array}

\usepackage{etoolbox}
\usepackage{setspace}
\newcounter{rowcntr}[table]
\renewcommand{\therowcntr}{\thetable.\arabic{rowcntr}}

\newcolumntype{N}{>{\refstepcounter{rowcntr}\therowcntr}c}

\AtBeginEnvironment{tabular}{\setcounter{rowcntr}{0}}

\title{Comparative study on charging-up of single, double and triple Gas Electron Multipliers (GEM)}

\author[a,b,1]{Vishal Kumar,\note{Corresponding author.}}
\author[a,b]{Supratik Mukhopadhyay,}
\author[a,b]{Nayana Majumdar,}
\author[a,b]{and Sandip Sarkar}

\affiliation[a]{Applied Nuclear Physics Division, Saha Institute of Nuclear Physics, Kolkata - 700064, India}
\affiliation[b]{Homi Bhabha National Institute, BARC Training School Complex, Anushaktinagar, Mumbai, Maharashtra-400094, India}

\emailAdd{vishal.kumar@saha.ac.in}

\abstract{In this paper, a detailed investigation has been carried out to understand the physics behind GEM charging-up and its effects on gain. Experiments have been performed on both double and triple GEM with the help of $^{55}$Fe X-ray source and a comparative study of these configurations along with the single GEM results observed in our previous work has been reported.
The increase in gain due to polarization of GEM foil dielectric and reduction in gain due to charge accumulation on dielectric are studied for various field configurations and different radiation intensities.
}

\keywords{Single, Double and Triple GEM detector, charging-up, charging-down, gain, radiation rate, dielectric polarization.}

\begin{document}
\maketitle
\flushbottom

\section{Introduction}
\label{S1}

The Gas Electron Multiplier (GEM) detectors \cite{sauli2016gas} are advanced Micro Pattern Gas Detectors (MPGDs) that are well known and used for their high position resolution, rate and discharge handling capability. These are considered to be suitable for large scale use in various experiments \cite{COMPASS, CMS1, CMS2, ALICE}. The detectors consist of GEM foils as multipliers, within which the multiplication of electrons takes place. The GEM foils are made up of a polyimide film of 50 $\mu m$ with 5 $\mu m$ copper cladding. Biconical holes in a hexagonal pattern are etched on the foil with the help of the double-mask photolithography technique. A standard GEM foil consists of holes with inner and outer diameters of 50 and 70 $\mu m$, respectively. These specifications can be altered according to requirements. Some examples of other GEM foils are single-mask GEM with hole asymmetry \cite{Aashaq2019} and thick GEM \cite{Alexeev2015} with hole and thickness of the order of hundreds of micron. The detector made up of standard GEM foils has been used in the present work.

Charging-up effects \cite{azmoun2006} are well known for their ability to modify the field around the dielectric material present in the active gas volume.  The multiplication occurs in GEM holes within which a considerable amount of dielectric remains exposed to the gas volume. As a result, the charging-up plays a vital role in GEM detector functioning. A detailed analysis of charging-up and its effects on gain for a single GEM have been described in \cite{Kumar2021}, with an investigation of charging-up with various field and rate configurations. The work also includes gain measurement techniques and methodology for varying radiation rates with the help of a collimator while using an extended radiation source.

The studies carried out on a single GEM in \cite{Kumar2021} were necessary to understand the basic features of the complex process. However, in actual experiments, single GEM configurations are rarely used, if at all. Most of the applications are based on triple or higher GEM configurations using three or more GEM layers \cite{COMPASS, CMS1, ALICE, Purba2017}. The use of multiple GEM foils has several advantages, such as the possibility of achieving larger gain values, improvement in position resolution due to charge spread, reduction in discharge probability due to lower field configurations, etc. The extension of the single GEM observations to double and triple GEM has been conducted to enhance the understanding of various aspects of charging-up in multi-layer GEMs. The effects of charging-up have been studied by varying parameters like GEM potential and radiation rates. These observations are crucial because essential parameters such as efficiency, energy resolution and position resolution are likely to be affected by gain variation of the GEM detector.

\section{Experimental Setup} 
\label{S2}

\begin{figure}[htbp]
\centering
\includegraphics[width= 0.6\linewidth,keepaspectratio]{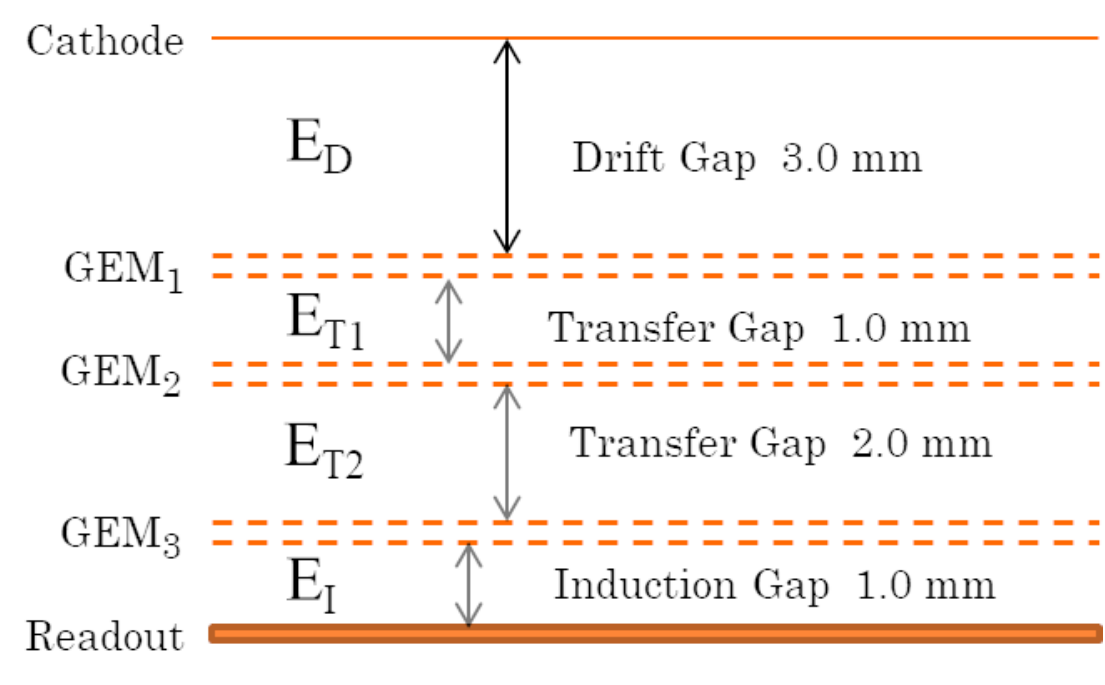}
\caption{\label{fig2a} Triple GEM schematic with drift field (E$_D$), transfer field (E$_{T1}$, E$_{T2}$) and induction field (E$_I$).}
\end{figure}

The GEM detector used here consists of a gas-filled volume having a cathode plane and a readout anode plane with one, or more, GEM foil(s) placed between them, dividing the detector gas volume into drift, transfer and induction regions, as shown in figure \ref{fig2a}.  The readout of the detector is made up of 256 strips running across both X and Y planes, from where 128 strips are shorted to a sum-up board to get a test pulse. These sum-up boards are as mentioned in~\cite{Patra2017}. The present studies have been performed with double and triple GEM setups, with 4.5, 2, 1.5 and 3, 1, 2, 1 mm drift to induction gaps, respectively. 
The negative potentials required by the detector have been provided with the help of a potential divider circuit. The voltage configurations used are listed in table \ref{table1}. The detector, along with the potential divider circuit, is placed in a copper box with Radio Frequency (RF) ground for a reduction in RF noise and exposure of X-ray while experimenting.

\begin{table*}[htbp]
\begin{center}
\caption{Experimental configuration} \label{table}
\scalebox{0.87}{
\begin{tabular}{ N c l l l l l l l l}
\hline
\multicolumn{1}{c}{Serial}& Description & E$_{D}$ & $\Delta$V$_{GEM1}$& E$_{T1}$ & $\Delta$V$_{GEM2}$ & E$_{T2}$& $\Delta$V$_{GEM3}$ & E$_{I}$ \\
\multicolumn{1}{c}{No.}& & (kV/cm) & (Volts) & (kV/cm)& (Volts) & (kV/cm) & (Volts) & (kV/cm) \\
\hline

\label{i} &S-GEM$_1$& 2.81 & 462 &  &  & & & 3.33 \\
\label{ii} &S-GEM$_2$& 2.81 & 508 &  &  & & & 3.33 \\
\label{iii} &D-GEM& 2.04 & 382 & 2.40 & 379 & & & 3.39 \\
\label{iv} & T-GEM$_1$ & 2.24&339&2.63&334&2.66&319&3.81  \\
\label{v} &T-GEM$_2$ & 2.30&348&2.69&342&2.72&327&3.90 \\ 
\label{vi}&T-GEM$_3$& 2.35&356&2.76&350&2.79&335&4.00 \\
\label{vii}& T-GEM$_4$& 2.44&369&2.86&363&2.89&347&4.14\\
\hline
\hline
\end{tabular}}
 \label{table} 
 \vspace{.1in}\\
Abbreviations for single, double and triple GEM detector are S, D and T-GEM respectively. 
\label{table1}
\end{center}
\end{table*}

Ar-CO$_2$ gas mixtures with volumetric ratios 74-26 and 80-20 have been used for single GEM and multi GEM detectors, respectively.
 In addition, the temperature, pressure and humidity of the room have been monitored with Bosch BME280 sensor~\cite{Bosch} while performing the experiment.

\section{Instrumentation and methodology}
\label{S3}
\begin{figure}[htbp]
\centering
\includegraphics[width= 0.6\linewidth,keepaspectratio]{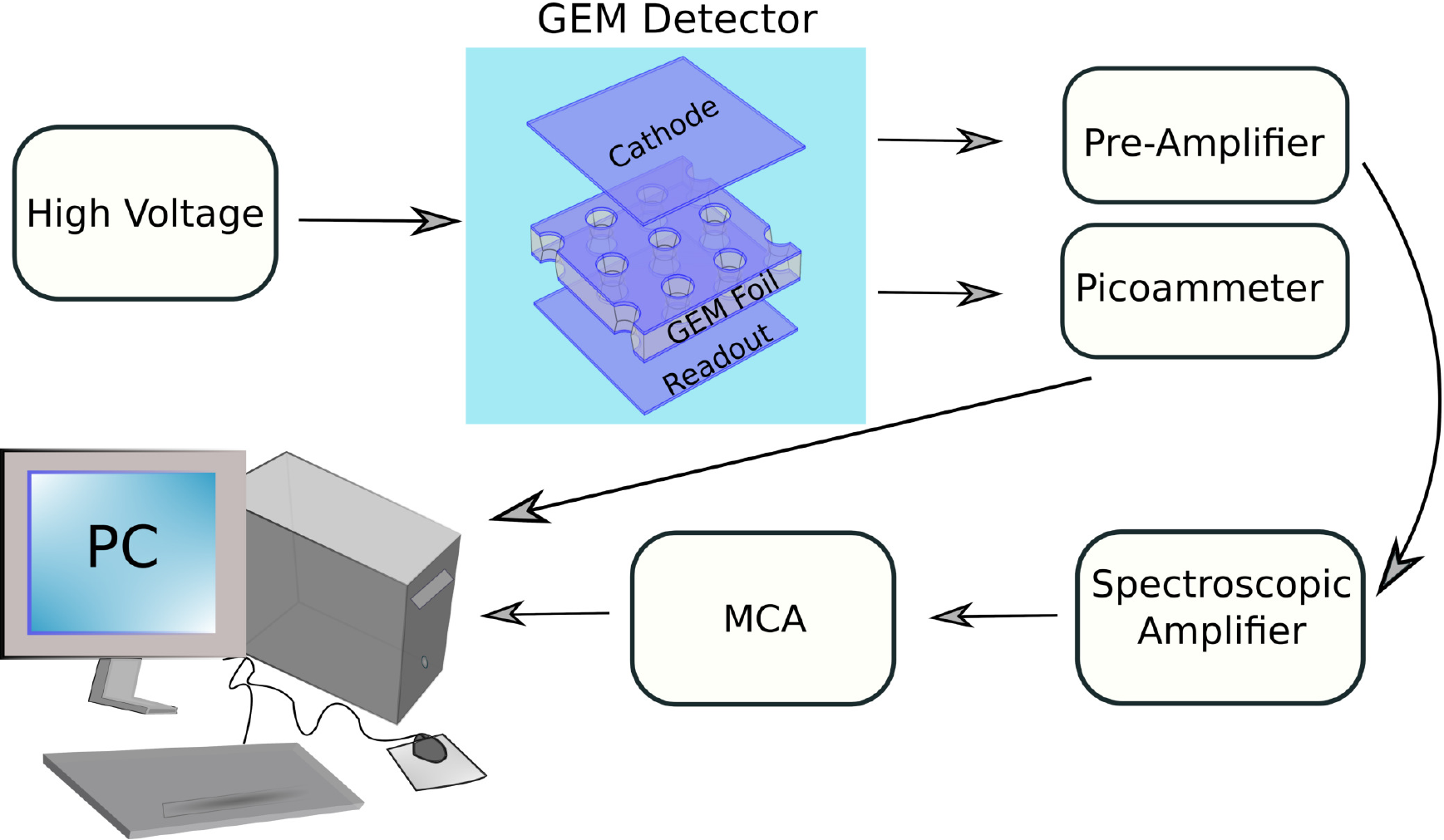}
\caption{\label{fig2} Schematic diagram of GEM detector setup.}
\end{figure}

 In order to perform the experiment, the signal from the sum-up boards mentioned in section~\ref{S2} is grounded with 120 k$\Omega$ resistors for impedance matching. For energy spectra measurement, a charge-sensitive preamplifier~\cite{caenA1422} is connected in parallel to one of the 120 k$\Omega$ resistors. The preamplifier integrates the signal and transfers it to a spectroscopy amplifier~\cite{Ortec672} for further amplification and shaping. The amplified Gaussian pulse from the spectroscopy amplifier is then sent to MCA~\cite{MCA} for collection of energy spectra data. For current measurement, a picoammeter~\cite{caenAH401D} is connected in series to the 120 k$\Omega$ resistors. 
The schematic diagram of the experimental setup is shown in figure~\ref{fig2}.
The current from the picoammeter and counts from the energy spectrum have been used to calibrate the MCA channels for gain. The gain obtained has been further processed to incorporate T/P corrections. The corrected gains in the following sections are the gain values obtained from the centroid of the Gaussian fit of the $^{55}$Fe 5.9 keV energy spectra with T/P corrections.
A detailed description of measurement technique and instrumentation for energy, current and T/P data acquisition can be found in our previous work on single GEM charging-up \cite{Kumar2021}.

\section{Radiation charging-up and charging-down}
\label{S4}
To understand the effect of radiation on the detector, it has been kept at respective potential for a day before performing the experiment. This is to ensure that the effects measured are solely due to radiation causing accumulation of charge on the GEM dielectric and not the polarization due to application of high field. The rate of the radioactive source has been varied by changing the source itself and by inserting a collimator in front of a distributed one.

\begin{figure}[htbp]
\centering
\includegraphics[width= 0.6\linewidth,keepaspectratio]{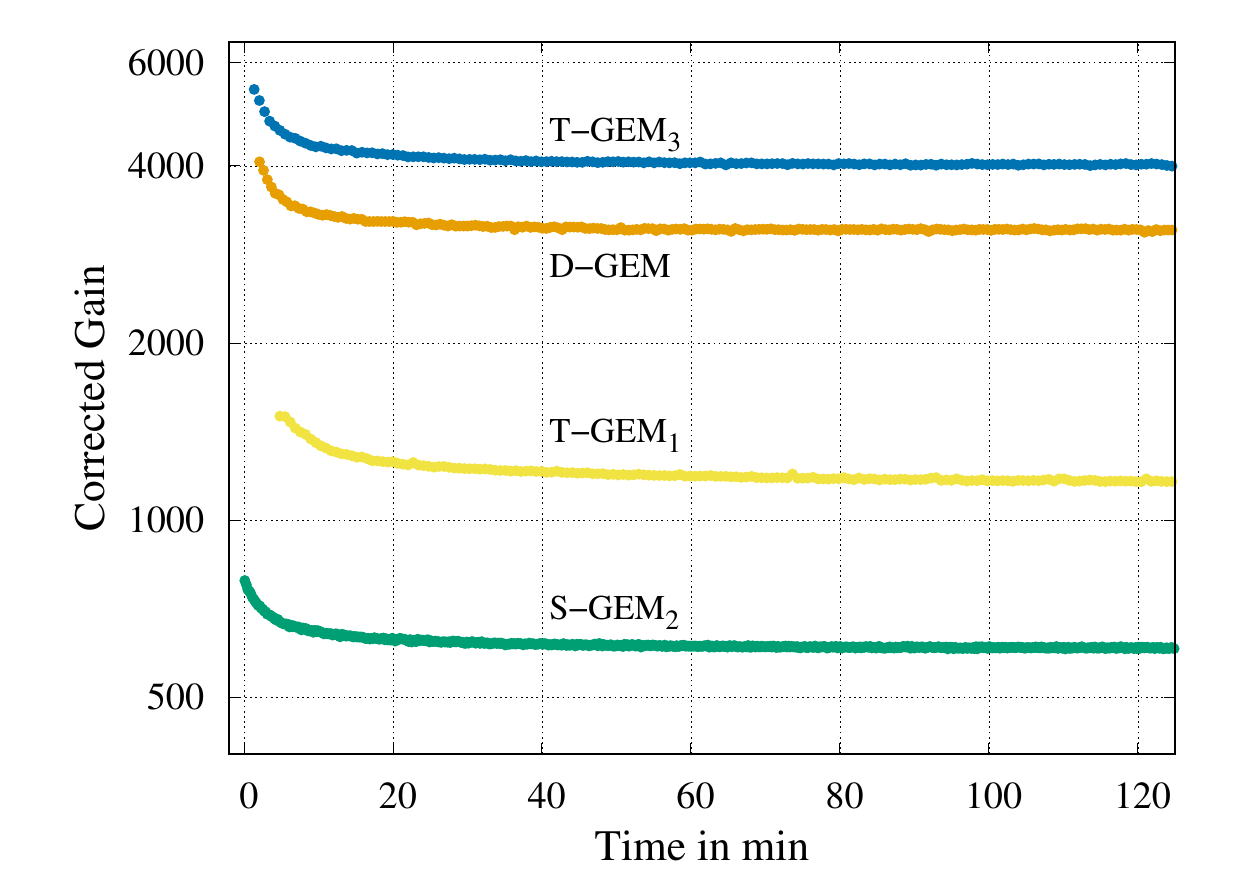}
\caption{\label{fig3a} Radiation charging-up of single, double and triple GEM with voltages configurations as mentioned in table \ref{table1}. A collimator of diameter 6 mm was used to control the rate of irradiation in all the cases.}
\end{figure}

As the source is placed on the detector at t=0 sec, the data taking starts and as shown in figure \ref{fig3a}, the gain value decreases and saturates to a steady value after around one or two hours, for all the configurations, namely, single, double and triple-GEM detectors. These measurements have been carried out for a fixed collimator and source configuration. However, the effective rates are slightly different due to different detector geometry parameters like the distance between cathode and source and the drift gap. Furthermore, the rate is lower in both double and triple GEM with respect to single GEM experiments due to the time difference in data taking. The half-life of $^{55}$Fe is 2.74 years and single GEM experiments had been conducted around one and a half years before (prior to the pandemic).

\begin{figure}[htbp]
\centering
\includegraphics[width= 0.6\linewidth,keepaspectratio]{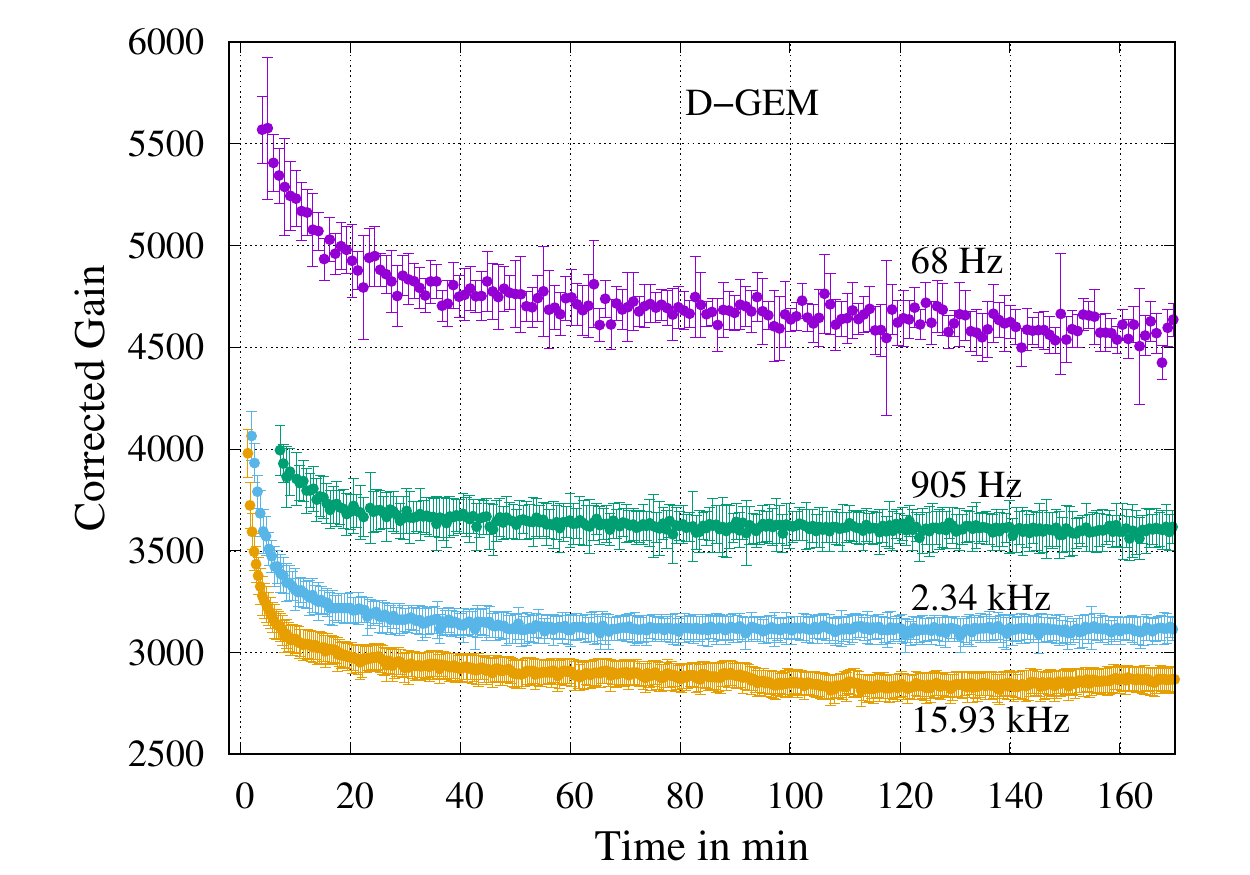}
\caption{\label{fig3b} Radiation charging-up of double GEM with various rates of irradiation.}
\end{figure}

Figure \ref{fig3b} shows the charging-up effects for the double GEM configuration for different irradiation rates. The gain decrease is faster for higher irradiation rates. The final equilibrium value of the gain is lower for a source of higher rate.

\begin{figure}[htbp]
\centering
\includegraphics[width= 0.6\linewidth,keepaspectratio]{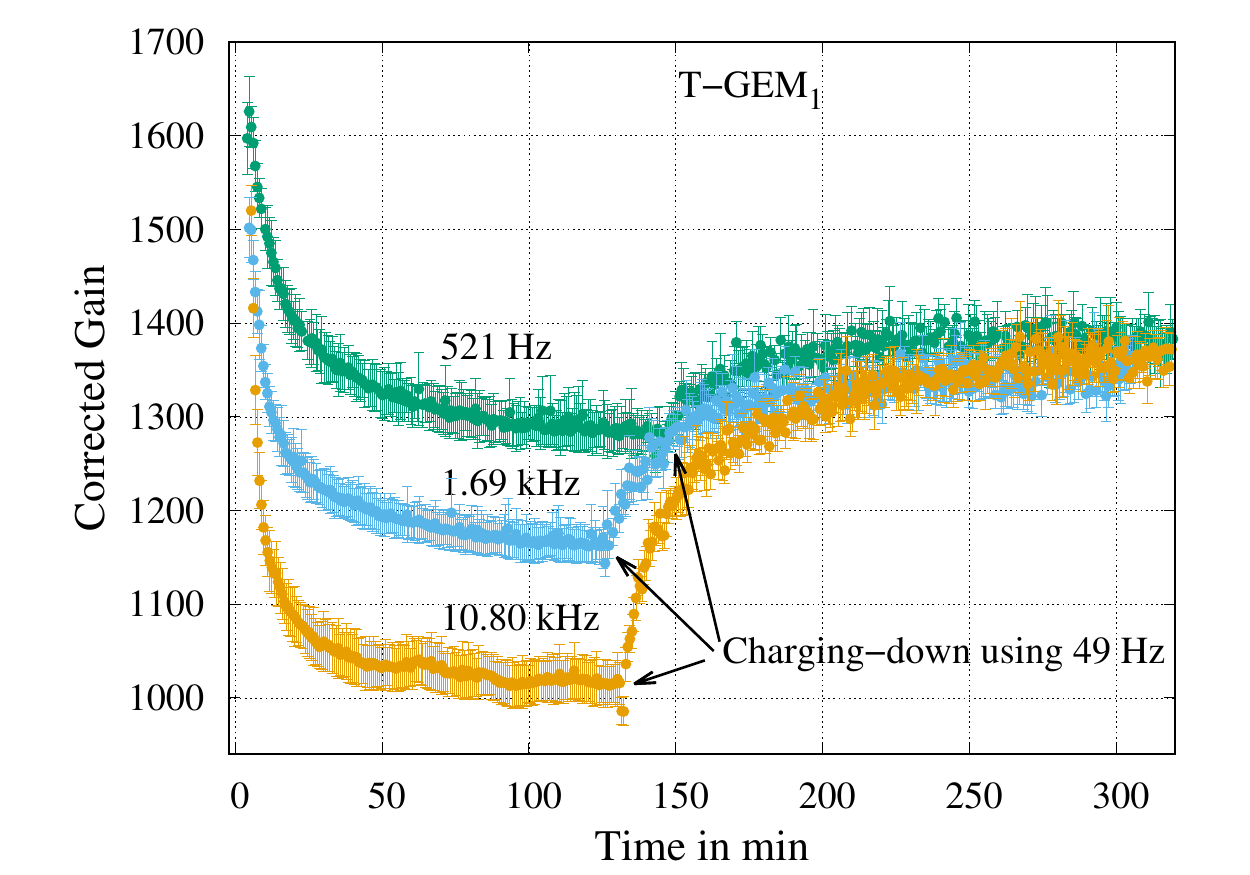}
\caption{\label{fig3c} Radiation charging-up and charging-down of triple GEM with rate variation and voltage configuration \ref{iv}.}
\end{figure}

In figure \ref{fig3c}, both charging-up and charging-down effects are presented for a triple GEM configuration. The effect of gain variations related to charging-up are similar to those observed for the double GEM configuration. The charging-down phenomenon, which has been investigated by replacing the original sources (0.521, 1.69. and 10.80 kHz) by a weak source (49 Hz) in comparison, keeping all other parameters fixed, shows that the gain increases and saturates at a higher value after several hours.

\begin{figure}[htbp]
\centering
\includegraphics[width= 0.6\linewidth,keepaspectratio]{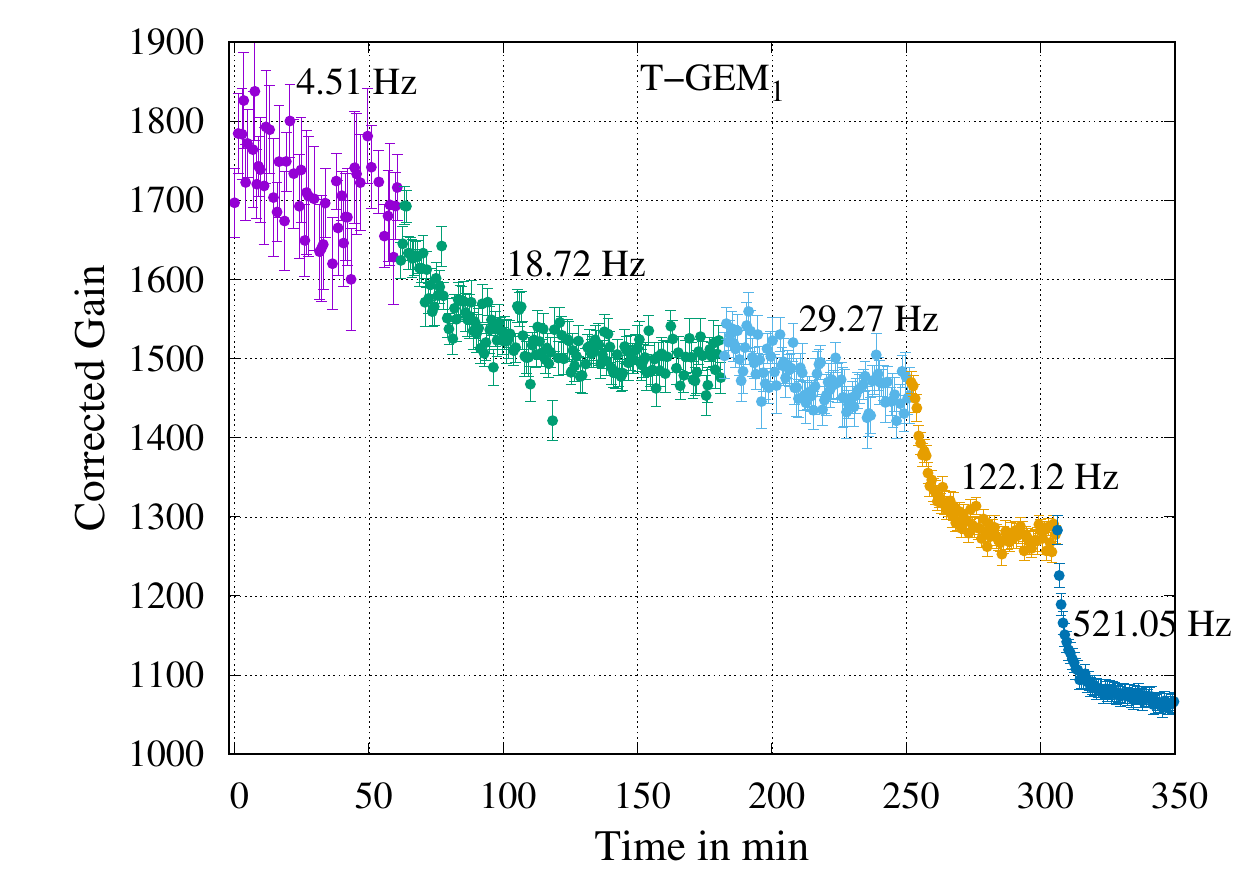}
\caption{\label{fig3d} Radiation charging-up of triple GEM with radiation rate controlled by aluminium foil.
 The experiment has been performed with voltage configuration \ref{iv}.}
\end{figure}

An alternative approach to modify the radiation rate is performed with one or more aluminium foil of thickness 50 $\mu m$ placed in between the detector and the radiation source. Here we have used two different $^{55}$Fe sources, both of which were first passed through a cylindrical collimator made up of stainless steel having a length of 13 mm and inner and outer diameter of 3 and 10 mm, respectively. This was to ensure the radiation beam coming out was parallel. The radiation rate is then modified by either placing or removing one or more aluminium foil in front of the collimator or by changing the source. Figure~\ref{fig3d} shows the reduction in gain value as the irradiation rate increases. The first two sets of data were obtained using the same source with and without 50 $\mu m$ foil, allowing 4.51 and 18.72 Hz irradiation rates, respectively. After around 180 minutes, the source was changed to a higher rate and data were obtained with foil thickness 100 $\mu m$, 50 $\mu m$ and without foil permitting irradiation rates of 29.27, 122.12 and 521.05 Hz, respectively.
Similar results have been observed in \cite{azmoun2006,Chernyshva2020} and what we have obtained before where the rate was varied by changing aperture size of the collimator. The rate obtained by passing through aluminium foils has been found in agreement with the calculations \cite{calculation}.

\begin{figure}[htbp]
\centering
\includegraphics[width= 0.6\linewidth,keepaspectratio]{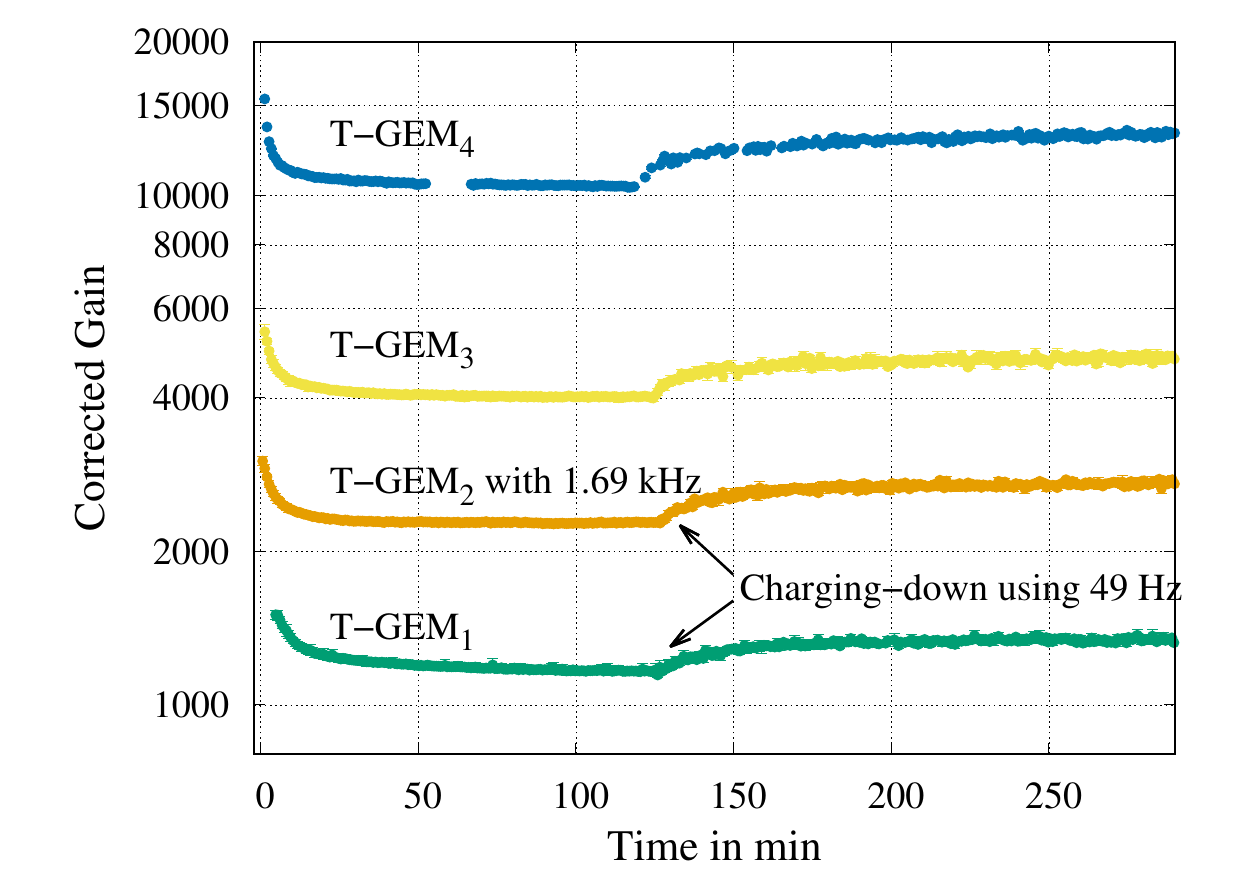}
\caption{\label{fig3e} Radiation charging-up and charging-down of triple GEM with voltage variation. For T-GEM$_4$ configuration the data taking was stopped for  a while, from around 55 to 65 minutes.}
\end{figure}

\begin{figure}[htbp]
\centering
\includegraphics[width= 0.6\linewidth,keepaspectratio]{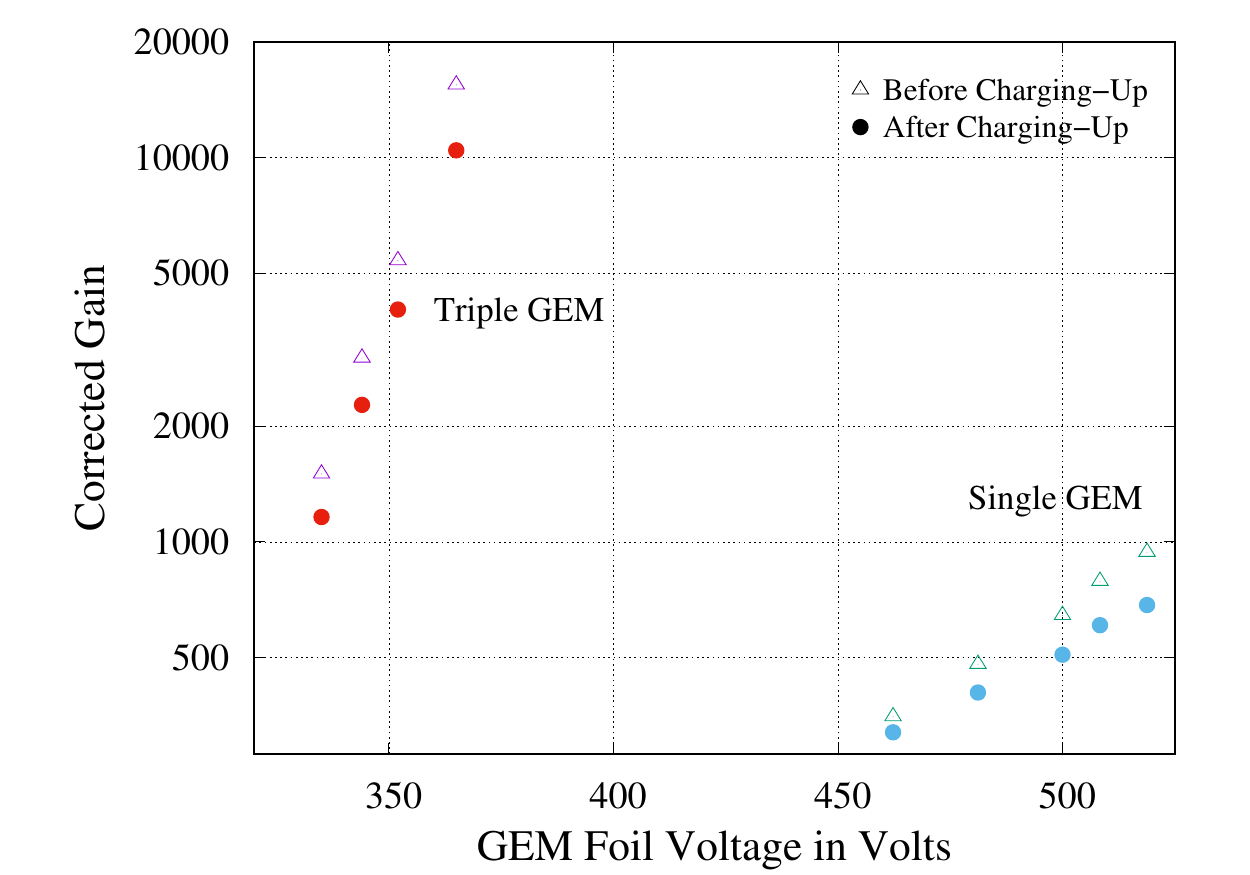}
\caption{\label{fig3f}Effect of GEM foil voltage on the gain of the detector. For triple GEM, the middle GEM foil voltage is considered since it is close to the average of all the three GEM foils.}
\end{figure}

The experiment of radiation charging-up has been performed with various voltage configurations in triple GEM with a 1.69 kHz radiation source. The charging-up in all the cases reduces the gain till it reaches a saturation level (see figure \ref{fig3e}). To measure the charging-down gain while the radiation source is removed, a weak 49 Hz source is placed to monitor the gain without causing a major impact on the detector gain.
The gain increases exponentially with GEM voltage for both single and triple GEM as shown in figure \ref{fig3f}. The gain data for both, before and after radiation charging-up have been plotted, exhibiting the reduction in gain due to radiation charging-up.

\section{Charging-up due to polarization of GEM dielectric}
\label{S5}
On application of high field on dielectric, it gets polarized. The polarization in general is a fast process that ranges in the order of $10^{-15}$ to $10^{-1}$ sec \cite{Kao2004}. However, the polarization in which we are interested in, is space charge polarization and it can last from minutes to hours \cite{Kao2004,Jonuz}.

\begin{figure}[htbp]
\centering
\includegraphics[width= 0.6\linewidth,keepaspectratio]{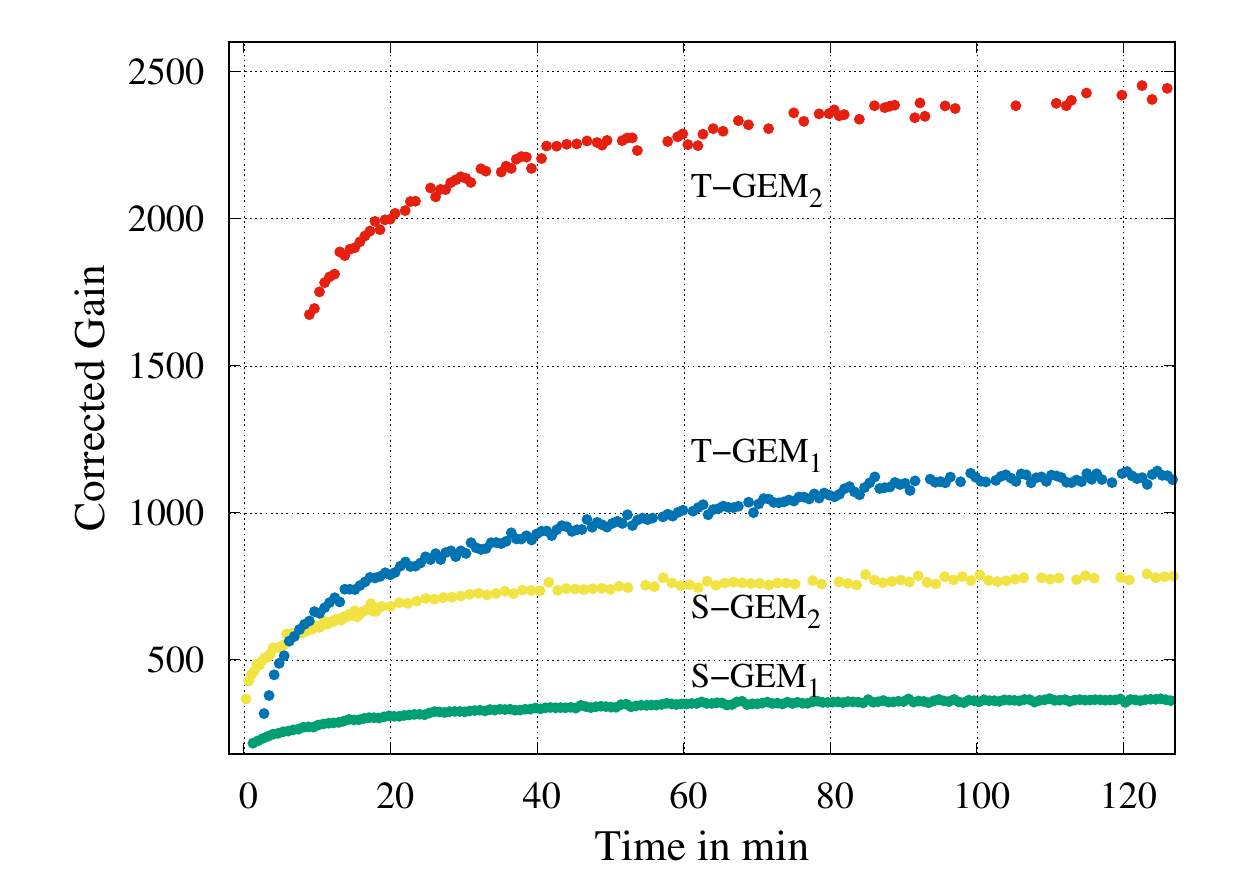}
\caption{\label{fig4a} Polarization charging-up with variation of applied voltage. The gain was estimated using irradiation at a very low rate (49 Hz for triple GEM and 120 Hz for single GEM).} 

\end{figure}

 To study space charge polarization, the detector is kept unbiased for 16 to 24 hours before experimenting. The potential has been applied with a ramp-up speed of 20V/sec. The data taking starts once the detector reaches the final values of potential. To minimize the effect of radiation charging-up, a low rate source has been used to get gain information. The gain is directly related to the modification of the field taking place inside the detector during the dielectric polarization process and helps in the understanding of the dielectric properties of the material.
As shown in figure \ref{fig4a}, the gain increases with dielectric polarization and reaches a saturation value, once it is complete. 
 The comparison of single and triple GEM data shows that the behaviour of polarization is similar along with the time required for saturation. This is because the polarization of dielectric is independent of the number of GEM foils in use.

\section{Summary and Conclusions}
\label{S6}
A comparative charging-up study on a double-masked GEM with single, double and triple GEM configurations has been performed. The experiments have been conducted with Ar-CO$_2$ gas mixture and with various irradiation rates using $^{55}$Fe soft X-ray sources. The radiation charging-up measurements have been performed by irradiating the detector kept on respective biases for days without irradiation. On the contrary, the polarization charging-up measurements have been performed on the detector which has been kept unbiased for several hours before the potential has been applied and a weak source has been kept to get the gain curve during polarization.

The radiation charging-up has been found to reduce the gain in all the configurations and to reach a saturation level within one or two hours. The saturation level depends upon the initial value of gain as well as on the radiation rate. In the case of charging-down, the strong source is replaced with a weak source after reaching the saturation level and the gain starts increasing till it reaches a saturation point. The results in figure \ref{fig3c} show that the saturation level attained with the weak source is constant and is independent of the level from where the charging-down starts. This is because the charges accumulated with different sources attain equilibrium with radiation rates at their saturation level of charging-up. However, once they are removed, the equilibrium is lost until it reaches the saturation level of the weak source.

The space-charge polarization of GEM foil(s) causes an increase in gain. The gain increases more rapidly, as well as to a higher level, in triple GEM as compared to single, as observed in figure \ref{fig4a} (the gain variation in T-GEM$_1$ is larger than those observed in S-GEM$_1$ and S-GEM$_2$). This effect is expected, as the polarization increases the gain in a single GEM, in triple GEM, all GEM foils are getting polarized simultaneously, causing a more rapid and higher increase in total gain.

As the number of GEM layers is increased, the reduction in gain due to radiation charging up is stronger. However, the time needed to reach the saturated value remains mostly unaltered (figure \ref{fig3a}). The increase in gain due to polarization charging is faster and larger for detectors with more layers (figure \ref{fig4a}). The results are in good agreement with the previous experimental works \cite{azmoun2006, Chernyshva2020} on radiation charging-up. The polarization charging-up of polyimide, discussed in \cite{Jonuz, Sessler}, also directs towards similar results.
To conclude, the effects of charging-up from both polarization and radiation are temporary and the gain modification decays rapidly with time and becomes negligible once the detector stays without bias voltage for hours.

\acknowledgments
The authors would like to thank Mr. Shaibal Saha for his technical help in assembling the GEM detectors. We would also like to acknowledge Dr. Purba Bhattacharya, Dr. Kanishka Rawat, Dr. Jaydeep Datta, Dr. Prasant K. Rout and Dr. Sridhar Tripathy for their valuable suggestions and advice in experimental work and data analysis. We are also grateful to Mr. Subendu Das, Ms. Promita Roy and Mr. Pralay Das for their precious time and help in the experimental work. 
We would also like to thank Prof. Chinmay Basu for his support. This work has partly been performed in the framework of the RD51 Collaboration and we would like to acknowledge the members of the RD51 Collaboration for their help and suggestions.


\begin{thebibliography}{99}


\bibitem{sauli2016gas}
F. Sauli, \emph{The Gas Electron Multiplier (GEM): Operating principles and applications}, \href{https://doi.org/10.1016/j.nima.2015.07.060}{\emph{Nucl. Instr. and Meth. in Phys. Res. A} {\bf 805} (2016) 2--24.}

\bibitem{COMPASS}
B. Ketzer {\textit{et al.}}, \emph{Performance of triple GEM tracking detectors in the COMPASS experiment}, \href{https://doi.org/10.1016/j.nima.2004.07.146}{\emph{Nucl. Instr. and Meth. in Phys. Res. A} {\bf 535~ 1-2} (2004) 314-318.}


\bibitem{CMS1}
D. Abbaneo {\textit{et al.}}, \emph{Upgrade of the CMS muon system with triple-GEM detectors}, \href{https://doi.org/10.1088/1748-0221/9/10/C10036}{\emph{Journal of Instrumentation} {\bf 9} C10036 (2014).}

\bibitem{CMS2}
G. Mocellin and on behalf of the CMS Muon Group, \emph{GEM detectors for the Upgrade of the CMS Muon Forward system},\href{https://iopscience.iop.org/article/10.1088/1742-6596/1390/1/012116/pdf}{ \emph{J. Phys.: Conf. Ser.} {\bf 1390} 012116 (2019).}

\bibitem{ALICE}
ALICE TPC collaboration  {\textit{et al.}}, \emph{The upgrade of the ALICE TPC with GEMs and continuous readout},\href{https://doi.org/10.1088/1748-0221/16/03/P03022}{ \emph{Journal of Instrumentation} {\bf 16} P03022 (2021).}

\bibitem{Aashaq2019}
A. Shah {\textit{et al.}}, \emph{Impact of single-mask hole asymmetry on the properties of GEM detectors}, \href{https://doi.org/10.1016/j.nima.2018.11.017}{\emph{Nucl. Instr. and Meth. in Phys. Res. A} {\bf 936} (2019) 459-461.}

\bibitem{Alexeev2015}
M. Alexeev {\textit{et al.}}, \emph{The gain in Thick GEM multipliers and its time-evolution}, \href{https://doi.org/10.1088/1748-0221/10/03/P03026}{\emph{Journal of Instrumentation} {\bf 10} P03026 (2015).}


\bibitem{azmoun2006}
B. Azmoun {\textit{et al.}}, \emph{A Study of Gain Stability and Charging Effects in GEM Foils},\href{https://doi.org/10.1109/NSSMIC.2006.353830}{ \emph{IEEE Nuclear Science Symposium Conference Record} (2006) 3847--3851.}

\bibitem{Kumar2021}
V. Kumar {\textit{et al.}}, \emph{Studies on charging-up of single Gas Electron Multiplier},\href{https://doi.org/10.1088/1748-0221/16/01/P01038}{ \emph{Journal of Instrumentation} {\bf 16} P01038 (2021).}

\bibitem{Purba2017}
P. Bhattacharya {\textit{et al.}}, \emph{3D simulation of electron and ion transmission of GEM-based detectors}, \href{https://doi.org/10.1016/j.nima.2017.06.054}{\emph{Nucl. Instr. and Meth. in Phys. Res. A} {\bf 870} (2017) 64-72.}

\bibitem{Patra2017}
R. N. Patra {\textit{et al.}}, \emph{Measurement of basic characteristics and gain uniformity of a triple GEM detector},\href{https://doi.org/10.1016/j.nima.2017.05.011}{ \emph{Nucl. Instr. and Meth. in Phys. Res. A} {\bf 862} (2017) 25-30.}

\bibitem{Bosch}
BME280 Datasheet - Bosch Global,\href{https://www.bosch-sensortec.com/media/boschsensortec/downloads/datasheets/bst-bme280-ds002.pdf}{ \emph{$www.ae-bst.resource.bosch.com/media/_tech/media/datasheets/BST-BME280-DS002.pdf$}, \emph{July 8} (2020).}





\bibitem{caenA1422}
CAEN A1422, \emph{Specification - $www.seltokphotonics.com/upload/iblock/713/7132eabc12ff687ab39e3c31871ea170.pdf$}, \href{https://www.seltokphotonics.com/upload/iblock/713/7132eabc12ff687ab39e3c31871ea170.pdf}{\emph{April 9} (2020).}

\bibitem{Ortec672}
ORTEC 672, \emph{Spectroscopy Amplifier - $www.physics.utah.edu/~springer/phys6719/experiments/manuals/672.pdf$}, \href{https://www.physics.utah.edu/~springer/phys6719/experiments/manuals/672.pdf}{\emph{April 9} (2020).}

\bibitem{MCA}
MCA8000D - Amptek, \emph{$www.amptek.com/-/media/ametekamptek/documents/products/mca-8000d-digital-multichannel-analyzer-specifications.pdf$},\href{https://www.amptek.com/products/multichannel-analyzers/mca-8000d-digital-multichannel-analyzer}{ \emph{July 8} (2020).}

\bibitem{caenAH401D}
CAEN AH401D, \emph{Online Manual - $www.caenels.com/wp-content/uploads/2015/04/AH401D_UsersManual_V1.5.pdf$},\href{https://www.caenels.com/wp-content/uploads/2015/04/AH401D_UsersManual_V1.5.pdf}{ \emph{April 9} (2020).}

\bibitem{Chernyshva2020}
M. Chernyshova {\textit{et al.}}, \emph{Effect of charging-up and regular usage on performance of the triple GEM detector to be employed for plasma radiation monitoring}, \href{https://doi.org/10.1016/j.fusengdes.2020.111755}{\emph{Fusion Engineering and Design}  {\bf 158} (2020) 111755.}

\bibitem{calculation}
G. Weber, X-Ray attenuation $\&$ absorption calculator, \emph{https://web-docs.gsi.de$\sim$stoe\_exp/web-programs/x\_ray\_absorption/index.php}, \href{https://web-docs.gsi.de/~stoe_exp/web_programs/x_ray_absorption/index.php}{\emph{Sep 28} (2021).}

\bibitem{Kao2004}
K. C. Kao, \emph{Dielectric phenomena in solids with emphasis on physical concepts of electronic processes},\href{https://www.sciencedirect.com/book/9780123965615/dielectric-phenomena-in-solids}{ \emph{Amsterdam: Elsevier Academic Press,} (2004).}

\bibitem{Jonuz}
Belma Alijagi\'{c} Jonuz, \emph{Dielectric Properties and Space Charge Dynamics of Polymeric High Voltage DC Insulating Materials}, \emph{Ph.D. thesis, Technische Universiteit Delft} (2007).

\bibitem{Sessler}
G.M. Sessler {\textit{et al.}}, \emph{Electrical conduction in polyimide films},\href{https://doi.org/10.1063/1.337646}{ \emph{J. Appl. Phys.} {\bf 60} (1986) 318--326.}

\end{thebibliography}
\end{document}